\documentclass[journal]{IEEEtran}
\IEEEoverridecommandlockouts
\usepackage{times,amsmath,color,amssymb,graphicx,epsfig,psfrag,algorithm,balance}
\usepackage{amsfonts,pifont,enumerate,cases}
\usepackage{mathrsfs} 
\usepackage[table]{xcolor} 
\usepackage{verbatim} 
\usepackage{bm}
\usepackage{cuted,stfloats}
\usepackage{algorithm}
\usepackage{algorithmic}
\usepackage{newtxmath}
\usepackage{longtable}
\usepackage{blindtext}
\usepackage{multirow}
\usepackage{float}
\usepackage{threeparttable}
\usepackage{makecell}
\usepackage[utf8]{inputenc}
\usepackage{url}
\usepackage{booktabs}
\usepackage{amssymb}
\usepackage{bbding}
\usepackage{pifont}
\usepackage{wasysym}
\usepackage{utfsym}

\usepackage{fontawesome}
\usepackage[algo2e,ruled,vlined,linesnumbered,lined,boxed,commentsnumbered]{algorithm2e}
\usepackage{amsmath,mathtools}
\usepackage[
    colorlinks=true,
    linkcolor=blue,
    citecolor=blue,
    urlcolor=magenta
]{hyperref}
\usepackage{array}

\begin{document}

\title{RIS-Assisted Secure ISAC: Fundamentals, Applications, and Future Directions}
\author{Zhendong Li, Jinyuan Huang, Weichun Zhao, Zhou Su, Enyu Shi, and Wen Chen
\thanks{Zhendong Li, Jinyuan Huang and Weichun Zhao are with the School of Information and Communication Engineering, Xi’an Jiaotong University, Xi’an 710049, China (e-mail: lizhendong@xjtu.edu.cn; 2246112066@stu.xjtu.edu.cn; zwc2856669274@stu.xjtu.edu.cn). Zhou Su is with the School of Cyber Science and Engineering, Xi'an Jiaotong University, Xi'an 710049, China (e-mail: zhousu@ieee.org). Enyu Shi is with the School of Electronics and Information Engineering, Beijing Jiaotong University, Beijing 100044, China (e-mail:  enyushi@bjtu.edu.cn). Wen Chen is with the School of Integrated Circuits, Shanghai Jiao Tong University, Shanghai 200240, China (e-mail:  wenchen@sjtu.edu.cn). (Corresponding author: Zhou Su)}
\vspace{-1.5em}
}
\maketitle

\begin{abstract}
Integrated sensing and communication (ISAC) is a key enabler for sixth-generation (6G) networks, yet its broadcast nature exposes transmissions to eavesdropping, especially under direct-link blockages. Passive reconfigurable intelligent surfaces (RISs) offer a promising physical-layer defense by reconstructing controllable virtual line-of-sight links. This article provides a design-oriented overview of passive RIS-assisted secure ISAC systems. We first outline hardware principles, cooperative architectures, and deployment topologies, followed by an analysis of synergetic gains, practical trade-offs, and representative 6G applications. We then detail key enabling technologies centered on a sensing-driven security loop, including cascaded channel estimation, decoupled active-passive beamforming, and predictive handover. A numerical case study under complete blockage quantifies the secrecy rate gains and signal-artificial noise decoupling achieved via sensing-assisted tracking. Finally, we highlight open challenges and future research directions toward practical deployment.
\end{abstract}

\section{Introduction}

\IEEEPARstart{A}{s} a transformative paradigm for sixth-generation (6G) wireless networks, integrated sensing and communication (ISAC) significantly improves spectral and hardware efficiency by unifying communication and radar sensing functionalities on a shared platform \cite{liu2022}. However, owing to the inherent broadcast nature of electromagnetic waves, dual-functional transmit beams are highly susceptible to interception while radiating energy into space for target detection. This dual exposure risk not only imperils the data confidentiality of legitimate users but also exposes sensitive environmental topologies and target state information to malicious eavesdroppers.

To mitigate these vulnerabilities, existing physical-layer security schemes for ISAC networks predominantly rely on active beamforming and artificial noise (AN) injection at the base station (BS) \cite{Su2024}. Although such transmitter-side precoding effectively balances secrecy rate and sensing accuracy under favorable line-of-sight (LoS) conditions, its defense capability depends heavily on a stable spatial propagation environment. In high-frequency bands and ultra-dense deployments, wireless links are severely prone to blockage by buildings and dynamic obstacles. Once the direct link is obstructed, conventional active countermeasures experience sharp performance degradation due to uncontrollable non-line-of-sight (NLoS) propagation. Consequently, overcoming physical channel blockage has become a critical bottleneck for deploying secure ISAC systems.

To tackle this limitation, the passive reconfigurable intelligent surface (RIS) offers a promising physical-layer defense paradigm through programmable propagation manipulation \cite{liu2023integrated}. Comprising an array of subwavelength reflecting elements, a passive RIS dynamically adjusts the phase shifts of incident waves to bypass physical barriers and establish controllable virtual LoS links among the BS, legitimate users, and sensing targets \cite{liz2022}. Compared with conventional active relays that entail substantial hardware costs, a passive RIS operates without active radio frequency (RF) chains or signal amplification. Consequently, it introduces negligible additive thermal noise and self-interference compared with active relays, which is vital for preventing fragile radar sensing echoes from being submerged in background noise.

Crucially, integrating passive RIS with secure ISAC extends well beyond conventional RIS-assisted secure communications by establishing a two-way sensing-security loop. Because a passive RIS lacks baseband processing capabilities, acquiring the channel state information (CSI) of non-cooperative or silent eavesdroppers via pilot training is fundamentally challenging. In secure ISAC, radar sensing echoes bridge this gap by extracting geometric and kinematic parameters (e.g., range, angle, and Doppler) to reconstruct wiretap channels and anticipate dynamic obstacles. In turn, synergistic active-passive beamforming and AN injection exploit these sensing priors to focus confidential signals toward legitimate users while steering jamming and sensing energy toward potential eavesdroppers, delivering notable radar signal-to-noise ratio (SNR) and secrecy rate gains \cite{chu2024}.

While recent surveys have comprehensively categorized RIS-aided physical-layer security algorithms and metasurface variants \cite{li2025}, they predominantly focus on static topologies or rely on idealized eavesdropper CSI assumptions. Against this backdrop, this article adopts a system-level design perspective tailored to severely blocked and dynamic environments, highlighting the closed-loop interplay between sensing-driven eavesdropper tracking/predictive handover and passive RIS virtual LoS reconstruction, supported by a complete numerical case study quantifying beam decoupling and sensing-secrecy trade-offs. Specifically, this article focuses on the system-level design of RIS-assisted secure ISAC: the enabling mechanisms, the trade-offs they impose, and the open problems that remain. The remainder of this article is organized as follows. Section \ref{sec:system} introduces the hardware principles and deployment topologies. Section \ref{sec:adv} analyzes the technical advantages of passive RIS integration and presents representative applications. Section \ref{sec:enab} details key enabling technologies, emphasizing cascaded channel estimation, joint active-passive beamforming, and sensing-driven predictive handovers. Section \ref{sec:case} presents a numerical case study validating secrecy gains and beam decoupling under direct-link blockages. Finally, Section \ref{sec:chal} identifies open challenges and future research directions, followed by concluding remarks in Section \ref{sec:con}.

\section{RIS-Assisted Secure ISAC Systems}
\label{sec:system}
In this section, we first introduce the hardware structure and electromagnetic principles of passive RIS, followed by an overview of RIS-assisted secure ISAC architectures and representative deployment schemes for effective security coverage.

\subsection{RIS Hardware and Electromagnetic Principles}
The technical superiority of a passive RIS stems from its high spatial degrees of freedom (DoF) in manipulating electromagnetic (EM) space. Structurally, an RIS is implemented as a metasurface array comprising a vast number of subwavelength programmable elements \cite{elmossallamy2020reconfigurable}. Each reflecting element typically consists of a metallic patch layer, a dielectric substrate, and a bottom fully reflective metallic ground plane. To achieve dynamic phase adjustability, low-power control components, such as positive-intrinsic-negative (PIN) diodes or varactor diodes, are integrated inside each element. Regulating the bias voltage of the control circuit changes the equivalent impedance characteristics of the element, and delivers precise control over the phase shift of incident signals. Throughout this article, we focus on quasi-passive reflection-type RIS (hereafter referred to as passive RIS). Unless specified otherwise, we assume ideal hardware conditions for conceptual illustration.

Unlike traditional active relays, a passive RIS adopts a controlled reflection mechanism without complex RF chain processing. Consequently, it introduces negligible additive thermal noise and self-interference compared with active relays, which is critical for extracting weak perception echoes in ISAC networks. By driving hundreds of reflecting elements with a specific phase shift mask, asymmetric deflection of EM waves or energy focusing can be realized to achieve precise channel control \cite{vaquero2026near}. This capability provides the foundational hardware substrate to bypass physical obstacles and overcome severe channel blockages at the physical layer.

\subsection{Architectures of RIS-Assisted Secure ISAC}
\begin{figure}[t]
    \centering
    \includegraphics[width=8cm]{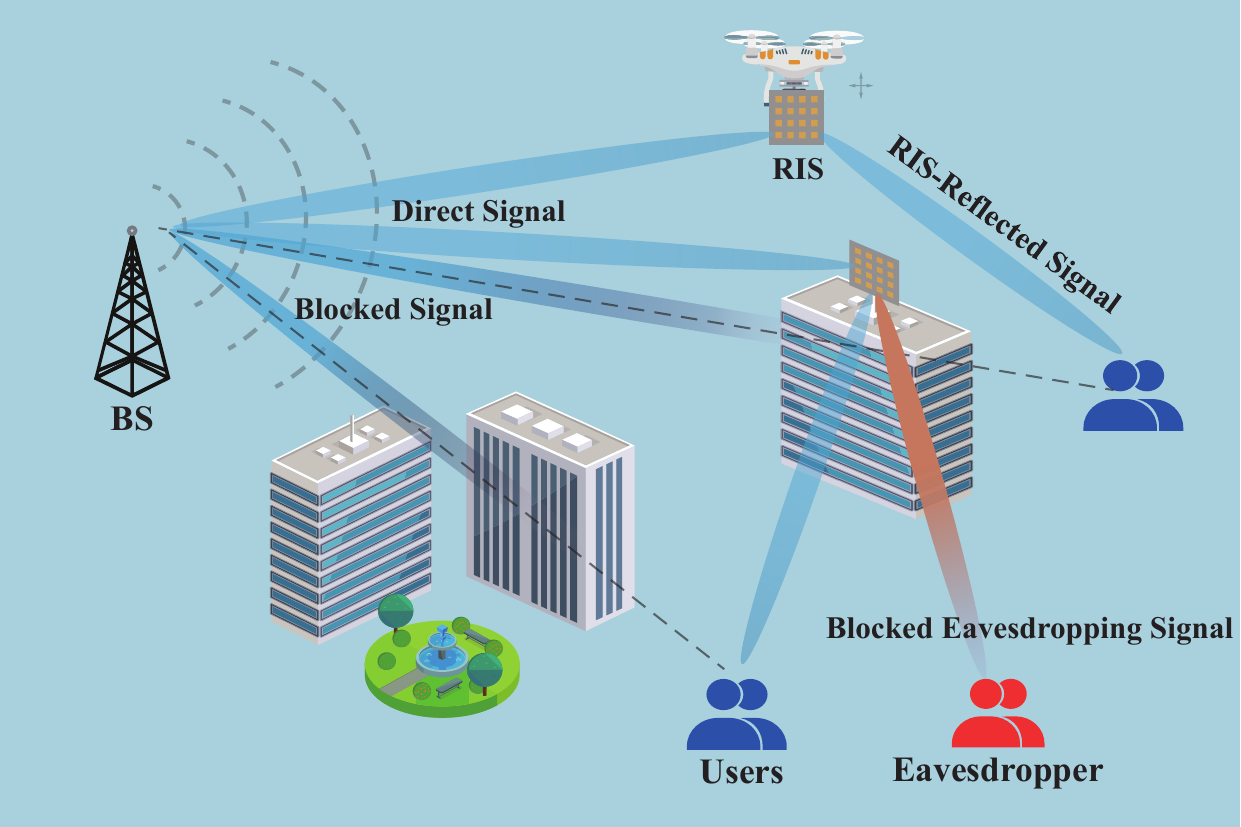}
    \caption{Architectures of RIS-assisted secure ISAC in the blocked environment.}
    \label{fig:1}
\end{figure}

Fig.~\ref{fig:1} presents the architectures of RIS-assisted secure ISAC in the blocked environment. This unified secure cooperative architecture comprises a BS, the passive RIS, legitimate users, and sensing targets. Within this architecture, the RIS serves as a spatial channel scheduling center, tasked with reconfiguring volatile channels to sustain the reliable operation of the secure ISAC system.

The operational principle relies on real-time reconstruction of cascaded channels. When interference or physical obstacles disrupt the direct path between the BS and the target, the system activates the RIS-assisted relaying procedure. The ISAC signals transmitted from the BS deviate from their original propagation routes and impinge on the RIS surface after the system explicitly identifies the RIS spatial location through sensing. The RIS then adopts reflective phase vectors optimized in advance to build cascaded virtual LoS links that bypass physical obstructions. This mechanism effectively avoids blockage of the direct path and restores secure communication services in NLoS regions. Meanwhile, highly directional reflective beams concentrate EM energy toward legitimate directions and naturally isolate potential eavesdropping paths through spatial configuration. This cooperative framework shifts the security defense paradigm from active transmit adaptation at the BS toward proactive physical channel topology reshaping.

In the sensing dimension, the RIS leverages beam focusing to compensate for the severe double-path loss that dampens echo signals over long distances, thereby extending the effective detection range of the system. Furthermore, the RIS provides extra spatial observation dimensions from diverse NLoS angles. Optimizing reflection phases improves target-parameter estimation for range, velocity, and azimuth, which not only maintains high-resolution perception under partial occlusion but also provides essential geometric priors for locating non-cooperative eavesdroppers.

\subsection{Deployment Schemes}
Flexible RIS deployment strategies further enhance the architecture to guarantee link robustness under complex physical environments.

From the perspective of spatial mobility, the architecture supports combined fixed and mobile deployment schemes. Deployment with fixed infrastructure places RISs on building facades, tunnel entrances, or street lamps. Leveraging the quasi-static property of such channels, this scheme determines optimal reflection positions through offline planning. It delivers persistent security compensation for long-term blind spots such as urban canyons. Its main technical advantage lies in low CSI acquisition overhead, which sustains stability of cascaded links with high energy efficiency. By contrast, mobile tracking deployment brings additional spatial DoF. Lightweight RIS units mounted on unmanned aerial vehicles (UAV) or ground mobile robots enable dynamic adjustment of reflection poses following target trajectories. Combined with intelligent scheduling algorithms including deep reinforcement learning, a mobile RIS adaptively optimizes its three-dimensional spatial position. It dynamically builds secure virtual LoS links according to requirements in real time for targets within highly volatile environments.

From the perspective of network topology, the deployment paradigm falls into centralized and distributed configurations. Centralized deployment places large-scale RIS arrays at key traffic hubs. It simplifies control protocols between the BS and RIS, and delivers broad secure coverage through high-gain narrow beams. Nevertheless, the distributed strategy offers better defensive resilience against extreme shadow fading brought by dense obstructions. Deploying multiple RIS units of small and medium sizes throughout the environment builds a multipath reflective network. These distributed units jointly reconfigure propagation links from different spatial angles. The approach mitigates connection outage risks triggered by single-point blockage. It also improves target-tracking accuracy and realizes a security upgrade from point-type coverage to domain-wide protection.

This evolution includes two key transitions. It moves from static patching to dynamic adaptive relaying, and shifts from gain obtained from a single array to multipath spatial diversity. These advances greatly extend the security boundaries of ISAC networks. With flexible deployment resource allocation, this architecture supports various vertical scenarios. Representative cases include vehicle coordination for intelligent transportation, illicit target tracking for UAV defense, and precision asset monitoring within the industrial Internet-of-things (IIoT).

\section{Advantages and Typical Applications}
\label{sec:adv}
The strategic value of the passive RIS-assisted secure ISAC architecture stems from physical topology reconfiguration. This paradigm delivers distinct gains in communication secrecy and sensing robustness, while offering a green, cost-effective security solution aligned with the requirements of 6G wireless networks.

\begin{table*}[b]
\centering
\caption{Qualitative Comparison of Different ISAC Paradigms under Blocked Environments (Indicative Ratings under Typical Deployment Assumptions)}
\label{tab:isac_comparison}
\resizebox{0.95\textwidth}{!}{
\begin{tabular}{lccccc}
\toprule
\textbf{ISAC Paradigm} & \textbf{Mobility Support} & \textbf{Power Consumption} & \textbf{Deployment Difficulty} & \textbf{Spectral Efficiency} & \textbf{Coverage Range} \\
\midrule
\textbf{Traditional Active ISAC} & Static Only & High & Low & Low (Blind spots) & Limited by LoS \\
\textbf{Active Relay-Assisted ISAC} & Static / Limited & Very High & High & Medium & Extended (Higher noise) \\
\textbf{Passive RIS-Assisted ISAC} & Fixed \& Mobile & Very Low & Low-Medium & High & Flexible Virtual LoS \\
\bottomrule
\end{tabular}
}
\end{table*}

\subsection{Secrecy and Sensing Synergetic Gains}
The core value of passive RIS lies in the comprehensive ISAC gains derived from its spatial DoF. Unlike conventional BSs fixed at static sites, passive RIS features superior deployment flexibility and morphological customizability, such as conformable skins on building facades, allowing it to embed deeply into traditional signal dead zones. When communication or detection links are severed by obstructions, real-time optimization of RIS reflection vectors constructs controlled virtual LoS paths. Such link reconfiguration restores physical connectivity, recovers secure transmission and target sensing, and turns adverse blocking environments into programmable security barriers \cite{shen2019secrecy}.

Physically, this gain originates from precise reshaping of spatial EM fields. Phase coordination across hundreds of elements produces highly directional narrow beams that weaken the correlation between legitimate and eavesdropping channels. With careful phase optimization, such spatial decoupling isolates eavesdropping paths at the physical layer to approach high secrecy capacity \cite{luo2021reconfigurable}. Meanwhile, RIS-assisted covert-communication schemes against unauthorized detectors have been explored in \cite{li2025covert}. In the sensing dimension, energy focusing partially compensates for the severe double-path loss suffered by long-distance perception echoes and supplies multi-angle spatial observations under NLoS conditions. Crucially, the resulting high-precision localization and environmental awareness feed back into the security layer to track non-cooperative eavesdroppers and guide adaptive beam nulling.

\subsection{Economic Advantages and Practical Trade-Offs}
For ultra-dense wireless networks, cost effectiveness and ecological sustainability govern large-scale commercialization. A passive RIS controls the EM environment without costly RF chains, power amplifiers, or digital baseband units. Unlike active relays that require high-capacity backhaul and continuous high power supply, a passive RIS introduces negligible additive thermal noise and self-interference compared with active relays \cite{huang2019reconfigurable}. This characteristic simplifies receiver design, prevents weak target sensing echoes from being overwhelmed by active-component noise, and lowers the hardware threshold for high-precision perception.

This low-power characteristic makes the architecture a promising candidate for green security solutions. Network operators can exploit the deployment flexibility of passive RIS to mitigate coverage holes with reduced capital and operational expenditures, even in energy-constrained or remote industrial environments. Nevertheless, realizing these benefits entails practical trade-offs: passive RIS links suffer from severe multiplicative (double-path) fading that necessitates large apertures, while the lack of baseband processing increases cascaded CSI acquisition overhead and makes performance sensitive to discrete phase-quantization errors. Table~\ref{tab:isac_comparison} provides an indicative qualitative comparison summarizing these performance and economic trade-offs across different ISAC paradigms.

\subsection{Application Scenarios}
\begin{figure*}[t]
    \centering
    \includegraphics[width=16cm]{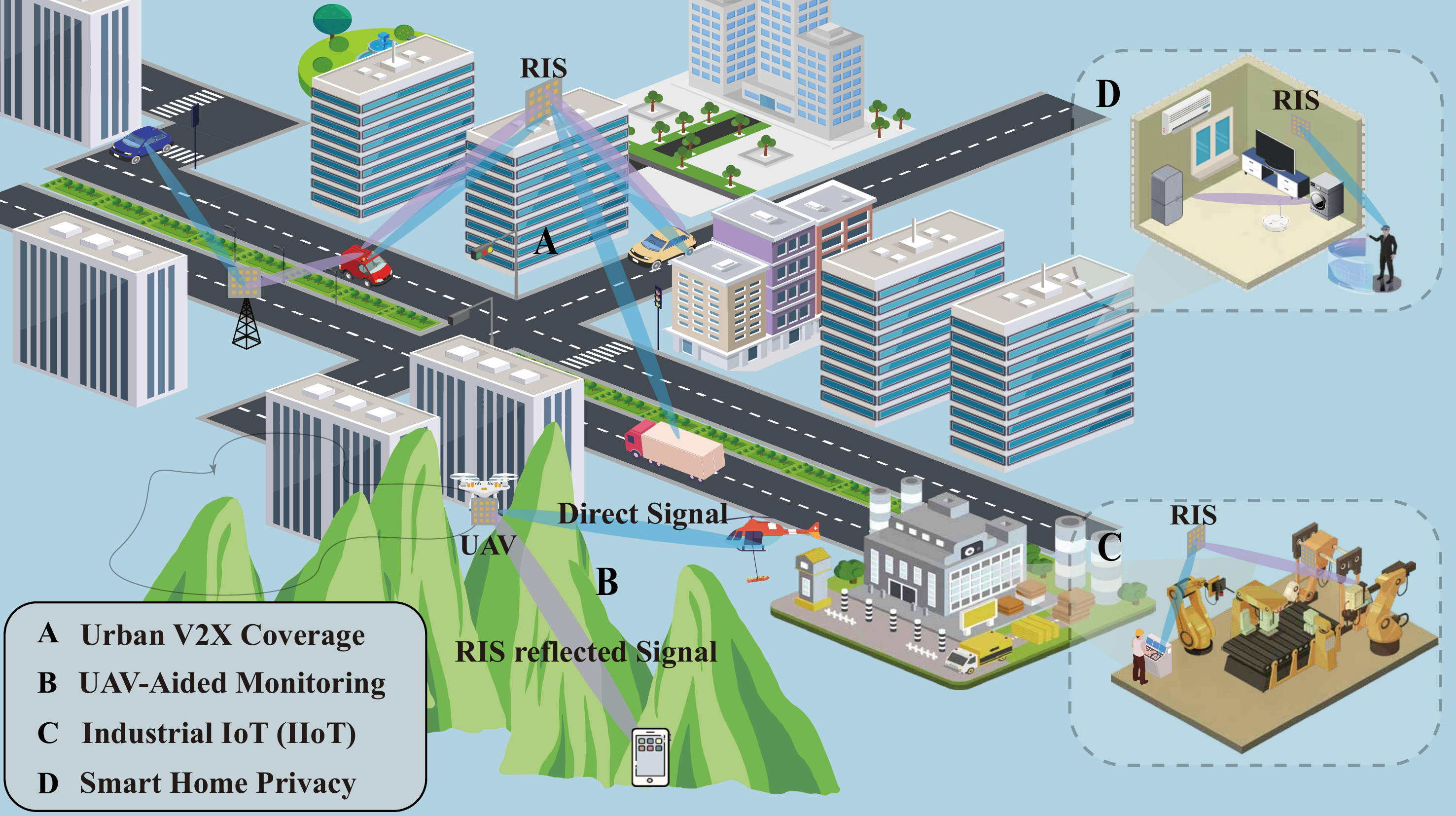}
    \caption{Typical applications of RIS-assisted secure ISAC.}
    \label{fig:use}
\end{figure*}

Differentiated deployment and network topology strategies enable this architecture to address distinct security threats and sensing demands across 6G vertical industries \cite{liu2023integrated}, as illustrated in Fig.~\ref{fig:use}.

\subsubsection{Urban V2X Coverage}
Under a fixed deployment paradigm, large centralized RIS arrays at major intersections and smaller distributed units at minor corners deliver secure coverage for vehicle-to-everything (V2X) communications. As shown in Scenario~A of Fig.~\ref{fig:use}, dense high-rise buildings within urban canyons create severe NLoS blind spots where malicious roadside or vehicular receivers can intercept cooperative driving messages around corners. Here, sensing demands real-time estimation of vehicle poses and velocities despite corner occlusions. Deploying passive RISs on building facades reconstructs cascaded virtual LoS links, leveraging real-time kinematic sensing to steer narrow confidential beams onto moving vehicles while maintaining spatial nulls toward non-cooperative roadside nodes.

\subsubsection{UAV-Aided Monitoring}
Employing a mobile tracking strategy, lightweight passive RIS units mounted on UAVs exploit three-dimensional spatial DoF to reconstruct directional virtual LoS links within NLoS regions. As shown in Scenario~B of Fig.~\ref{fig:use}, mountain ridges and forest canopies break direct links during low-altitude search-and-rescue or emergency monitoring, exposing sensitive telemetry and video feeds to unauthorized ground or aerial eavesdroppers. This scenario requires continuous 3D target tracking across volatile NLoS terrain. Drawing upon real-time situational awareness, a mobile RIS tunes its reflection geometry along target trajectories to sustain tracking continuity while isolating wiretap paths at the physical layer.

\subsubsection{Industrial IoT}
Combining fixed and distributed configurations in IIoT environments, small passive RIS units mounted on workshop surfaces establish a multipath reflective network that turns physical barriers into programmable security walls. Illustrated in Scenario~C of Fig.~\ref{fig:use}, modern smart factories contain dense metallic machinery and moving robotic arms that induce abrupt link blockages, raising risks of unauthorized interception of proprietary control commands and topology leakage. Sensing requirements center on high-precision, low-latency positioning of collaborative robots amid severe metallic clutter. Passive RIS networks bypass these metallic barriers, confine confidential control signals within designated work cells, and enhance multi-angle localization accuracy.

\subsubsection{Smart Home Privacy}
Under a fixed and distributed layout, compact passive RIS units embedded within walls or furniture bypass indoor partitions to achieve blind-zone-free coverage. As shown in Scenario~D of Fig.~\ref{fig:use}, indoor concrete walls and appliances block direct links between gateways and residents, while broadcast sensing echoes create severe privacy threats such as through-wall eavesdropping on vital signs and behavioral patterns. The sensing objective is fine-grained micro-motion and health monitoring. Directional beam enhancement and physical-layer obfuscation confine sensing energy strictly within legitimate indoor rooms, protecting both perception echoes and private data against unauthorized interception.

\section{Key Enabling Technologies}
\label{sec:enab}
Translating the physical gains of passive RIS into system-level security protection capabilities requires overcoming key technical bottlenecks, including sensing-assisted cascaded channel acquisition, decoupled active-passive beamforming, and predictive handover under dynamic blockages.

\subsection{Cascaded Channel Estimation}
Within NLoS scenarios with physical occlusions, acquiring reliable CSI acts as the logical starting point for building physical-layer security boundaries. Because a passive RIS lacks active baseband processing capabilities, cascaded channel estimation faces high dimensionality and low received SNR. In secure ISAC networks, CSI acquisition must explicitly distinguish between cooperative and non-cooperative nodes. For legitimate users, the BS-RIS and RIS-user cascaded channels can be estimated via dedicated pilot sequences exploiting RIS spatial sparsity to reduce training overhead \cite{zhou2022channel}. 

By contrast, non-cooperative eavesdroppers do not transmit pilots, rendering conventional training infeasible. When a potential eavesdropper also acts as a sensing target, ISAC radar echoes bridge this gap by extracting high-precision geometric and kinematic parameters, including range, angle, and Doppler frequency, to backward infer relative positions of moving obstacles and eavesdroppers, thereby reconstructing geometry-based virtual LoS wiretap channels. Nevertheless, for fully passive eavesdroppers with weak radar cross-sections buried in clutter, echo-based channel reconstruction remains fundamentally limited and requires dedicated clutter-suppression and target-classification mechanisms.

\subsection{Active-Passive Beamforming}
Joint active and passive beamforming is essential to balance communication secrecy and radar sensing accuracy. In a passive RIS-assisted secure ISAC system, BS transmit signals must simultaneously satisfy user quality-of-service (QoS) constraints, target illumination requirements, and eavesdropper suppression goals. To respect the single phase-profile constraint of passive elements while supporting dual-functional operation, the RIS reconfiguration can be decoupled between the secure downlink and the sensing illumination, i.e., two distinct phase profiles are optimized instead of one shared profile.

By jointly coordinating the BS transmit precoding matrix, AN injection, and the RIS phase-shift masks, the system achieves fine-grained spatial energy decoupling. Specifically, when the sensing target is also a potential eavesdropper, confidential information beams focus directional gain toward legitimate users while synthesizing deep spatial nulls at the eavesdropper direction \cite{liu2023drl}. Concurrently, AN waveforms form nulls toward legitimate users to avoid degrading their reception while concentrating energy on the eavesdropper/target, simultaneously jamming malicious interception and guaranteeing sufficient echo SNR to approach the Cramér-Rao bound or tracking accuracy thresholds. Dynamic weighting factors further adapt these metrics to operational priorities, mitigating resource competition between secrecy and sensing.

\subsection{Dynamic Tracking and Handover}
\begin{figure}[t]
    \centering
    \includegraphics[width=0.95\linewidth]{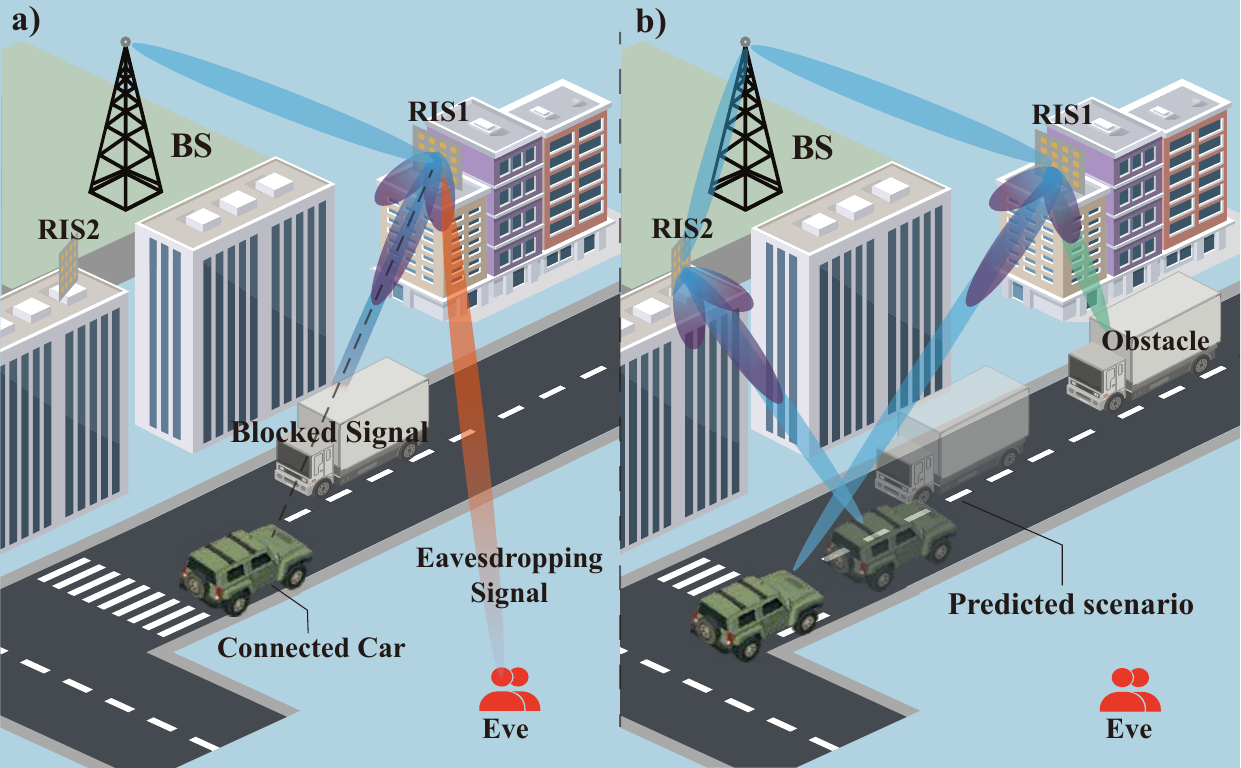}
    \caption{Comparison of handover schemes: (a) Reactive handover causing link disruption and information leakage; (b) Proposed sensing-driven predictive handover enabling seamless security via real-time obstacle tracking.}
    \label{fig:predictive_handover}
\end{figure}

To counter dynamic channel blockages caused by moving obstacles, traditional reactive handovers usually lead to instantaneous security link interruptions and defense failures, as illustrated in Fig.~\ref{fig:predictive_handover}(a). The core technique of this architecture lies in a sensing-driven predictive adaptive scheduling mechanism, depicted in Fig.~\ref{fig:predictive_handover}(b). Leveraging radar echoes scattered from dynamic obstacles and mobile eavesdroppers, the BS extracts kinematic parameters such as velocity vectors, azimuth angles, and Doppler frequencies. Building upon recursive state estimators such as the extended Kalman filter (EKF) or long short-term memory networks \cite{xia2024sensing}, the BS extrapolates relative spatial-geometric motion trajectories for obstacles, eavesdroppers, and controlled virtual LoS links over upcoming time slots.

Based on this trajectory prediction, a proactive handover trigger is established: whenever a geometric collision model predicts that the current virtual LoS link's received SNR or secrecy margin will violate a preset threshold within a short look-ahead window, the BS pre-computes and pre-loads the optimal reflection phase mask onto an unblocked standby RIS before actual link disruption occurs (Fig.~\ref{fig:predictive_handover}(b)). This predictive mechanism supports seamless beam transitions under favorable conditions and substantially reduces privacy-leakage and identity-exposure risks stemming from transient link outages and frequent reconnection signaling, achieving full-lifecycle adaptive defense under dynamic topology evolution.

\section{Case Study}
\label{sec:case}
To evaluate the security and sensing performance of the passive RIS-assisted ISAC system under severe physical blockages, numerical simulations are conducted in a representative multi-user scenario. The passive RIS serves as a core reconfigurable node to establish cascaded virtual LoS links when the direct transmission paths between the BS and users or targets are completely blocked. To satisfy the single phase-profile constraint of passive elements while supporting dual-functional operation, the RIS is configured with two distinct phase profiles, one governing the forward secure transmission and one governing the reflected sensing echoes. Meanwhile, a mobile eavesdropper simultaneously acts as the sensing target, whose state is recursively tracked from the received echoes via an EKF to reconstruct the geometry-based wiretap channel. The primary objective is to maximize the total secrecy rate while satisfying BS transmit power limits, EKF tracking accuracy constraints, and unit-modulus RIS phase constraints. Because the BS transmit precoding vectors, the AN vector, and the two RIS phase profiles are tightly coupled, an alternating optimization framework is utilized to solve the resulting nonconvex problem.

In the simulation setup, a dual-functional radar-communication (DFRC) BS serves $K =3$ legitimate users along with one mobile eavesdropper that also acts as the sensing target, positioned in close physical proximity to the user cluster. The passive RIS consists of $M = 36$ reflecting elements with an inter-element spacing of $\lambda/2$, matching the antenna layout at the BS. The operating carrier frequency is set to $f_c = 30\text{ GHz}$, and the maximum transmit power of the BS is $P_{\text{max}} = 35\text{ dBm}$. All cascaded channels associated with the RIS follow a distance-dependent path loss model, with a reference path loss of $C_0 = -20\text{ dB}$ at a reference distance of $D_0 = 1\text{ m}$. The path loss exponents are configured as $2.0$ for the BS-RIS link and $2.2$ for the RIS-user links, with Rician factors of $0\text{ dB}$ and $8\text{ dB}$, respectively, whereas the RIS-eavesdropper link is modeled as a geometry-based LoS channel reconstructed from EKF tracking. The background noise powers at the legitimate users, the eavesdropper, and the BS receiver are set to $-90\text{ dBm}$, $-85\text{ dBm}$, and $-100\text{ dBm}$, respectively.

\begin{figure}[t]
    \centering
    \includegraphics[width=0.8\linewidth]{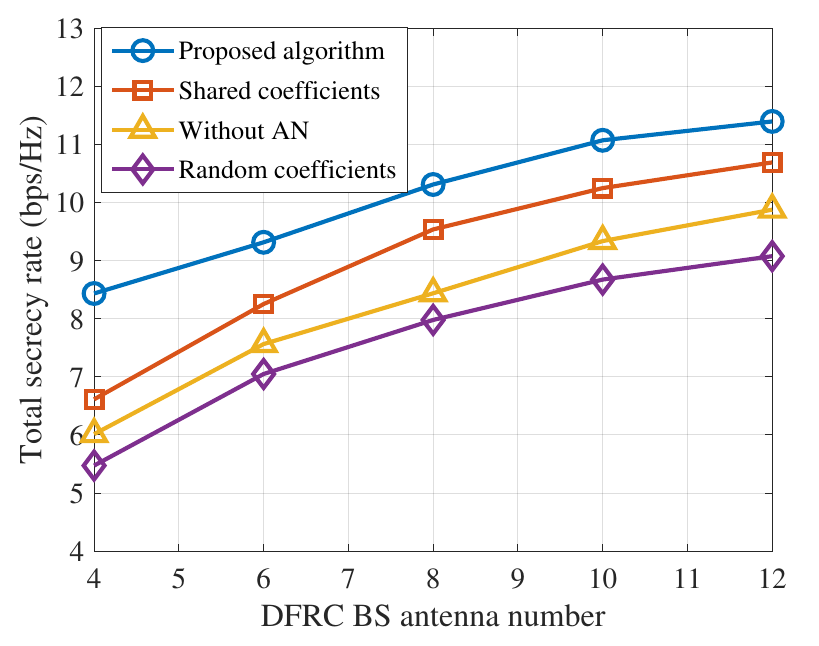}
    \caption{Total secrecy rate versus the number of DFRC BS antennas under different benchmarks.}
    \label{fig:ris_elements_rate}
\end{figure}

As shown in Fig.~\ref{fig:ris_elements_rate}, the total secrecy rate increases monotonically with the number of DFRC BS antennas across all configurations, confirming the expanded active spatial DoF for multi-user beam separation and wiretap suppression. The proposed joint optimization algorithm consistently outperforms all baseline configurations across all antenna scales. The performance disparities highlight the critical role of coordinated active-passive design and AN injection. Specifically, the benchmark enforcing identical RIS reflection coefficients for forward transmission and echo sensing (Shared coefficients) limits spatial flexibility, while adopting uncoordinated random phase shifts for echo sensing (Random coefficients) fails to collect target echoes coherently. Furthermore, the benchmark disabling AN injection (Without AN) exhibits a substantial rate drop, demonstrating that dedicated AN is indispensable for safeguarding confidential information under sensing constraints.

\begin{figure}[t]
    \centering
    \includegraphics[width=0.85\linewidth]{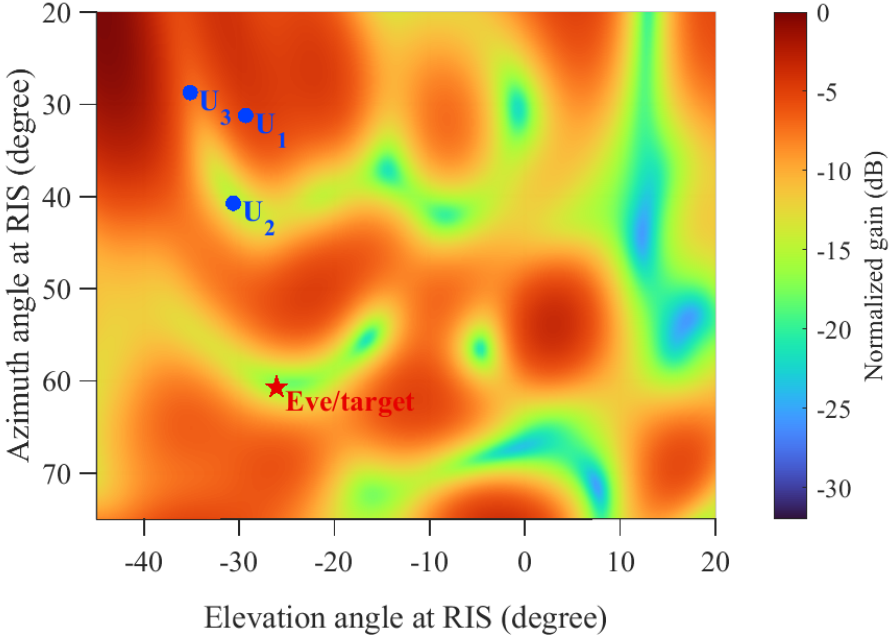}\\[-2pt]
    {\footnotesize (a) Confidential ISAC signal.}\\[8pt]
    \includegraphics[width=0.85\linewidth]{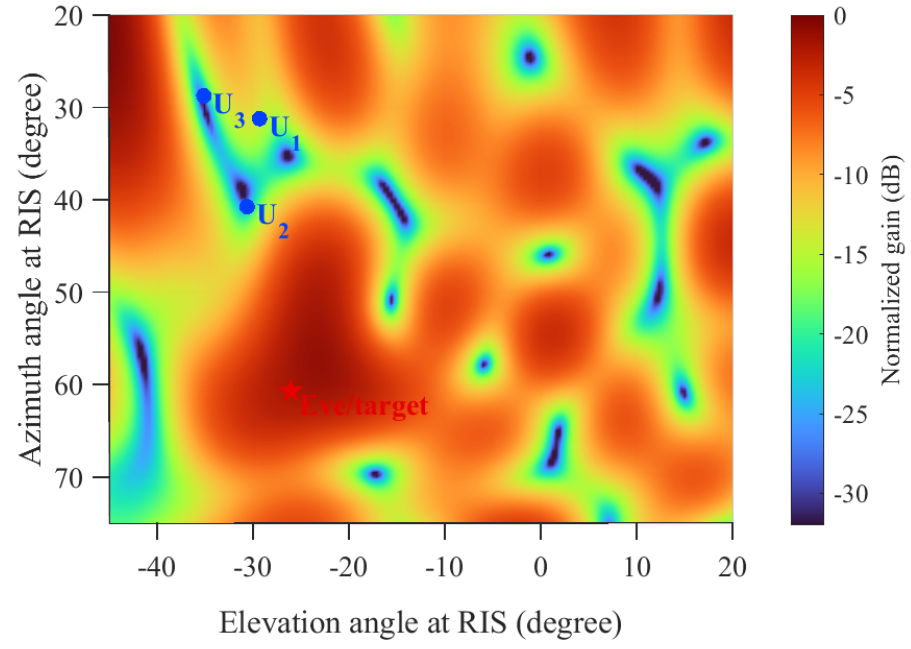}\\[-2pt]
    {\footnotesize (b) AN signal.}
    \caption{Normalized spatial gain (beampattern) at the RIS.}
    \label{fig:beampattern}
\end{figure}

Fig.~\ref{fig:beampattern} illustrates the normalized spatial beampatterns synthesized at the RIS for the two signal components. As shown in Fig.~\ref{fig:beampattern}(a), the confidential ISAC waveform concentrates directional gain over the legitimate user cluster $(U_1, U_2, U_3)$; because the eavesdropper also serves as the sensing target, the same beam also illuminates the eavesdropper/target region, which is where the radar acquires its echo. Conversely, Fig.~\ref{fig:beampattern}(b) shows the AN waveform, which places deep spatial nulls at $(U_1, U_2, U_3)$ to preserve legitimate reception while concentrating jamming energy on the eavesdropper/target region. Together, the two components realise the division of labour underlying secure ISAC: the ISAC waveform carries the data and provides the illumination, while the AN waveform protects the data by suppressing reception at the eavesdropper.

In summary, by providing additional spatial DoF and adaptively reshaping physical wireless channels, the passive RIS-assisted secure ISAC system effectively alleviates the resource competition bottleneck between communication and sensing. It enables high-precision beam alignment, directional security enhancement, and effective physical-layer interference suppression even under severe channel blockage.

\section{Open Challenges and Future Directions}
\label{sec:chal}
Translating the enabling technologies in Section~\ref{sec:enab} into practical 6G deployments still faces key bottlenecks across synchronization, covert threat detection, and real-time orchestration.

\subsection{Synchronization and Hardware Limits}
High-precision time-frequency synchronization underpins passive RIS-assisted secure ISAC. Without active RF chains or baseband units, a passive RIS cannot directly estimate its own clock offsets or phase noise. Under dynamic blockage, severe double-path fading yields extremely low pilot SNR, hindering blind synchronization convergence. Furthermore, practical hardware impairments, including discrete phase quantization, amplitude-phase coupling, and thermal bias drifts, distort reflection beams, shift spatial nulls, and elevate sidelobes, simultaneously degrading radar resolution and leaking confidential energy. Developing robust passive synchronization protocols and hardware-aware beamforming algorithms therefore remains a critical bottleneck.

\subsection{Hidden Eavesdropper Defense}
Most physical-layer security schemes assume active non-cooperative eavesdroppers whose channels can be inferred from emitted signals. In practical 6G verticals such as smart homes and IIoT, however, malicious receivers often remain completely silent. Under such extreme conditions, the system must rely entirely on radar sensing to detect abnormal entities. In cluttered NLoS environments, weak echoes from hidden eavesdroppers easily blend into background clutter. Exploiting the spatial DoF of passive RIS to extract discriminative radar signatures, such as micro-Doppler and radar cross-section features, for target classification and adaptive beam nulling remains a demanding challenge for closing the sensing-driven security loop.

\subsection{AI-Driven Security Management}
Fast-varying topologies require millisecond-level responsiveness for RIS reconfiguration and predictive handover. Traditional convex optimization methods, such as semidefinite relaxation and block coordinate descent, involve high-dimensional matrix operations whose complexity typically scales cubically with the number of RIS elements, failing to meet real-time latency constraints. Integrating artificial intelligence (AI) for autonomous security management is thus an inevitable trend. Nevertheless, deep learning and reinforcement learning face strong out-of-distribution generalization demands under abrupt blockages, while joint secrecy-tracking optimization involves high-dimensional hybrid action spaces (continuous precoding and discrete RIS phases) prone to training divergence. Developing lightweight, physics-informed AI frameworks for real-time closed-loop security orchestration represents a core future direction.

\section{Conclusion}
\label{sec:con}
This article presented a comprehensive overview of passive RIS-assisted secure ISAC from the perspectives of fundamentals, applications, enabling technologies, and future directions. By establishing controllable virtual LoS links, a passive RIS provides additional spatial DoF to enhance physical-layer security and sensing robustness when direct links are blocked or severely degraded. The case study demonstrated that combining sensing-driven tracking with joint active-passive beamforming and decoupled signal-AN energy allocation substantially improves secrecy rates while ensuring reliable target illumination under complete direct-link blockage. More importantly, passive RIS-assisted secure ISAC extends the role of metasurfaces beyond link enhancement toward a closed-loop, sensing-driven control of the EM environment. Realizing this vision, however, calls for progress in passive synchronization and hardware-aware design, hidden eavesdropper detection, and lightweight AI-driven security orchestration under dynamic network conditions. Addressing these challenges will be critical to transforming passive RIS-assisted secure ISAC into a practical foundation for secure and resilient 6G networks.
\balance
\bibliographystyle{IEEEtran}
\bibliography{reference}

\end{document}